\documentclass[%
 amsmath,amssymb, longbibliography,preprint,aip]{revtex4-1}

\usepackage{dcolumn}
\usepackage{bm}
\usepackage[mathlines]{lineno}
\usepackage{siunitx}
\usepackage{amsmath}

\usepackage{mathptmx}
\usepackage{etoolbox}
\usepackage[table]{xcolor}

\usepackage{float}
\usepackage{graphicx}
\usepackage{color}

\usepackage{amsthm}

\usepackage[cal=cm,scr=euler]{mathalpha}

\begin{document} 
\title{Relativistic and quantum fluid dynamics: nonlinear momentum evolution}
\author{Fei Wang}

\email{fei.wang@kit.edu; ORCID: 0000-0003-3318-3264}
\affiliation{Institute of Nanotechnology (INT),
Karlsruhe Institute of Technology (KIT),\\
Hermann-von-Helmholtz-Platz 1, 76344 Eggenstein-Leopoldshafen, Germany}

 \date{\today}

\begin{abstract}
We present an alternative framework for nonlinear momentum evolution in fluid dynamics, in which relativistic effects are proposed as an underlying mechanism for nonlinear behavior, even in the low-velocity regime. By interpreting the second-order velocity diffusion equation from a quantum-mechanical perspective, we establish a connection between higher-order viscous transport and the reduced Planck constant, identifying the second-order viscosity coefficient with \(\hbar\). Furthermore, we show that the coupling of second-order velocity diffusion and nonlinear momentum evolution leads to a Dirac-type equation for quantum fluids. These findings reveal a potential connection among relativistic effects, higher-order fluid transport, and quantum wave dynamics.

\end{abstract}

\maketitle

\section{Introduction}

In the first part of this research series~\cite{wang2026density}, we proposed a kinetic Gibbs--Duhem relation to describe linear momentum evolution in fluid dynamics:
\begin{equation}
    \mathbf{u}\cdot \mathrm{d}(m\mathbf{u})+\mathrm{d}P=0,
    \label{eq:1}
\end{equation}
where $m$ is the mass of a fluid particle, $\mathbf{u}$ is its velocity, and $P$ is a pressure-like Lagrange multiplier associated with the particle mass and volume. The corresponding energy is defined as
\begin{equation}
    E=m\mathbf{u}^{2}+P.
\end{equation}
Integrating the kinetic Gibbs--Duhem relation [Eq.~\eqref{eq:1}] gives $P=-m\mathbf{u}^{2}+C$, where $C$ is an integration constant, yielding $E=C$. Within the proposed framework, this construction provides a \textit{Galilean-invariant} energy quantity. Furthermore, the kinetic Gibbs--Duhem leads to the linear momentum evolution equation
\begin{equation}
    \frac{\mathrm{d}(m\mathbf{u})}{\mathrm{d}t}=-\nabla P.
\end{equation}

In the second part of this research series~\cite{wang2026thermodynamics}, the kinetic Gibbs--Duhem framework was extended to describe the coupling between fluid dynamics and electromagnetic fields. A generalized Lagrange multiplier, $P=P_1+P_2$, was introduced to account for the fluid kinetic energy and electromagnetic energy, respectively, through
\begin{equation}
    m\mathbf{u}^{2}+P_1=C_1,\qquad
    \varepsilon\mathbf{E}^{2}+P_2=C_2.
\end{equation}
Here, $C_1$ and $C_2$ are two constants, $\varepsilon$ is the electrical permittivity and $\mathbf E$ denotes
 the electrical field.
Within this framework, the Maxwell equations were equivalent to the linear momentum evolution of the coupled fluid--electromagnetic system.

Building upon these studies, the present work extends the kinetic Gibbs--Duhem framework to incorporate relativistic and quantum effects. In particular, we investigate the physical origin and evolution of nonlinear momentum contributions, which were not considered in our previous studies. The central question is: How do nonlinear momentum evolution and higher-order velocity diffusion emerge from relativistic fluid dynamics, and how are they connected to quantum mechanics?

To address this question, we develop a theoretical framework for fluid dynamics in both low- and high-velocity regimes. We show how relativistic effects contribute to nonlinear momentum evolution, including in the low-velocity limit. Furthermore, we establish a quantum-mechanical interpretation of the second-order velocity diffusion equation and demonstrate how its coupling with nonlinear momentum evolution leads to the Schr\"{o}dinger and Dirac equations in the corresponding quantum-fluid regimes. These results provide a theoretical connection between relativistic fluid dynamics, higher-order viscous transport, and quantum wave mechanics.

The remainder of this paper is organized as follows. Section II presents the proposed fluid dynamics theory in the low- and high-velocity regimes. Section III discusses its physical interpretation and validity. Section IV summarizes the main results.

\section{Theory}

We consider a relativistic fluid particle with mass $m$. 
Its energy can be written as~\cite{landau1975classical,salmon2026similarities}
\begin{equation}
 E_0=mc^2=\gamma m_0 c^2,\ \gamma=\frac{1}{\sqrt{1-\mathbf{u}^2 /c^2}},
 \end{equation}
where $c$ is the speed of light in vacuum, $\gamma$
is the Lorentz factor,  $\mathbf{u}$ is the relative motion velocity of the particle, and 
$m_0$ is the rest mass.
The three-dimensional velocity is defined as 
\begin{equation}
 \mathbf{u}=\frac{\mathrm d \mathbf r}{\mathrm d t}=\sum_i^3 \dot{x_i}\mathbf e_i=\sum_i^3 u_i \mathbf e_i,
\end{equation}
where $\mathbf{r}$ denotes the particle's position vector, and $u_i$ is the component of its velocity along the $i$-th coordinate direction, defined by the unit vector $\mathbf{e}_i$.

\subsection{Low speed limit}

For $|\mathbf{u}|\ll c$, a Taylor expansion of the energy $E_0$ leads to 
\begin{equation}
 E_0=m_0c^2+\frac{1}{2}m_0\mathbf u^2+ \frac{3}{8}\frac{m_0}{c^2}\mathbf u^4+\cdots,
\end{equation}
The first term, $m_0c^2$ is a constant background energy which is not included in the following derivation for brevity.
The other terms are variables in term of the velocity $\mathbf u$.
As these velocity-dependent terms are non-objective with respect to Galilei transformation, 
we modify the energy by adding Lagrange multiplier terms:  
\begin{align}
 E=E_1+E_2+\cdots=\boxed{m_1 \mathbf u^2+ P_1} +\boxed{m_2\mathbf u^4+ P_2}+\cdots 
  =\sum_{n=1}^\infty (m_n \mathbf u^{2n}+P_n),\ n \in\mathbb{Z}^{+}.
\end{align}
The Lagrange multiplier $P_n$ is associated with the mass $m_n=\chi_n m_0/c^{2n-2}$,
$\chi_n=(2n)!/4^n(n!)^2$ is the binomial coefficient of the Taylor expansion.
When defining a mass density $\rho_n=m_n/\delta V_n$,
the Lagrange multiplier $P_n=p_n\delta V_n$ corresponds to the  constraint for the volume $\delta V_n$ of the particle, 
where $p_n$ is the pressure in the classic thermodynamic interpretation~\cite{blokhuis1995young,gibbsCollectedWorksThermodynamics1928}.

The total derivative of the modified energy $E$ reads 
\begin{equation}
 \mathrm d E=\boxed{m_1\mathbf{u}\cdot \mathrm d\mathbf{u}+ [\mathbf{u}\cdot \mathrm d(m_1\mathbf{u})+\mathrm dP_1]}+\boxed{m_2\mathbf{u}^2 \mathrm d\mathbf{u}^2+ [\mathbf{u}^2 \mathrm d(m_2\mathbf{u}^2)+\mathrm dP_2]} +\cdots,
\end{equation}

We first examine the leading terms in the first boxed expression.
Applying the dissipation principle~\cite{wang2026density,wang2023thermodynamically}, the first term gives rise to a diffusion equation for the velocity:
\begin{equation}
 \frac{\mathrm d \mathbf u}{\mathrm d t}=D_1\nabla^2(m_1 \mathbf{u}),
\end{equation}
where $D_1$ is a positive velocity diffusivity that is related to the viscosity governing macroscopic fluid motion (see Section III). 
The corresponding rate of energy dissipation is
\begin{equation}
 \frac{\mathrm d (m_1 \mathbf{u}^2)}{\mathrm d t}=-D_1 m_1^2\nabla \mathbf{u}:\nabla \mathbf{u}\leq 0,
\end{equation}
which is consistent with the energy dissipation property of the Navier--Stokes equations~\cite{yue2020thermodynamically,hoffman2021energy}.
The rest terms in the first box are a kinetic Gibbs-Duhem relation
\begin{equation}
\mathbf{u}\cdot \mathrm d(m_1\mathbf{u})+\mathrm dP_1=0.
\label{eq:7}
\end{equation}
Dividing both sides of Eq.~\eqref{eq:7} by $\mathrm d x_i$, noting that $\mathrm d \mathbf r/\mathrm d x_i=\mathbf{e}_i$, 
and summing all the equations result in the linear momentum evolution
\begin{equation}
 \frac{\mathrm d  (m_1\mathbf u)}{\mathrm d t}=-\nabla P_1.
 \label{eq:12}
\end{equation}
Integrating Eq.~\eqref{eq:7} gives rise to an explicit expression for $P_1$ as 
\begin{equation}
 P_1=-m_1\mathbf{u}^2+C_1.
\end{equation}
This result indicates that the  modified system energy $E_1=m_1\mathbf{u}^2+P_1=C_1$ is Galilei invariant. 
Here, $C_1$ is an integration constant.
The physical meaning of $P_1$ is the pressure energy associated with the linear momentum $m_1\mathbf u$.
Equivalently, $p_1$ is the pressure associated with the linear momentum  density $\rho_1\mathbf u$.

Next, we examine the second boxed term in the total differential of the modified energy. Similarly, applying the dissipation principle to the term $(m_2\mathbf{u}^2)\mathrm{d}\mathbf{u}^2$ yields the following diffusion equation:
\begin{equation}
 \frac{\mathrm d \mathbf u^2}{\mathrm d t}=D_2\nabla^2(m_2 \mathbf{u}^2),
 \label{eq:15-9}
\end{equation}
where $D_2$ is a positive diffusion coefficient.
The resulting energy dissipation reads
\begin{equation}
 \frac{\mathrm d (m_2 \mathbf{u}^4)}{\mathrm d t}=-D_2 m_2^2(\nabla \mathbf{u}^2)^2\leq 0.
\end{equation}
The remaining terms in the second box likewise form a kinetic Gibbs--Duhem relation, ensuring the Galilean invariance of the energy:
\begin{equation}
\mathbf{u}^2 \mathrm d(m_2\mathbf{u}^2)+\mathrm d P_2=0.
\label{eq:11}
\end{equation}
Integrating  Eq.~\eqref{eq:11} gives rise to 
\begin{equation}
 P_2=-m_2\mathbf{u}^4+C_2.
\end{equation}
This result indicates that the modified system energy $E_2=m_2\mathbf{u}^4+P_2=C_2$ is Galilei-transformation invariant. 
Here, $C_2$ is an integration constant.
The physical meaning of $P_2$
is the pressure energy
associated with the second order non-linear momentum.

Dividing both sides of Eq.~\eqref{eq:11} by $\mathrm d x_i$ results in the second order momentum evolution
\begin{equation}
 \mathbf{u}\frac{\mathrm d  (m_2\mathbf u^2)}{\mathrm d t}=-\nabla P_2.
 \label{eq:19-9}
\end{equation}

A further analysis of the higher-order terms in the modified energy leads to a generalized velocity diffusion equation,
\begin{equation}
    \frac{\mathrm{d}\mathbf{u}^n}{\mathrm{d}t}
    =D_n\nabla^2(m_n\mathbf{u}^n),
\end{equation}
together with the corresponding generalized kinetic Gibbs--Duhem constraint,
\begin{equation}
    \mathbf{u}^{2n}\,\mathrm{d}(m_n\mathbf{u}^{2n})
    +\mathrm{d}P_n=0.
\end{equation}

\subsection{High speed fluid particle}

The preceding analysis applies to fluid particles in the low-velocity regime, where $|\mathbf{u}|\ll c$. For more general cases, the system energy is extended to the relativistic form
\begin{equation}
    E=\gamma m_0 \mathbf c^2+P,
\end{equation}
where  $P$ is a generalized Lagrange multiplier introduced to ensure the Lorentz invariance of the total system energy. 
In this formulation, the relativistic contribution is incorporated directly through the Lorentz factor, while the additional term $P$ accounts for the constraints required by the proposed framework. This approach differs from conventional relativistic fluid formulations, which typically introduce the four-velocity to describe the motion of fluid particles in a manifestly Lorentz-covariant manner~\cite{andersson2021relativistic,martimuller2003,romatschke2010}. Instead, the present approach incorporates relativistic effects through the modified energy and its associated constraint.

The total derivative of the energy reads
\begin{equation}
 \mathrm d E=\boxed{ \gamma m_0 \mathbf c\cdot  \mathrm d\mathbf c+  [\mathbf c\cdot  \mathrm d (\gamma m_0\mathbf c)+ \mathrm d P]}.
 \label{eq:20}
\end{equation}
The first term in Eq.~\eqref{eq:20} leads to a  diffusion equation with relativistic effect:
\begin{equation}
\frac{\mathrm d \mathbf{c}}{\mathrm d t}=\nabla ^2 (\gamma m_0 \mathbf{c}).
\end{equation}
As  $\mathbf{c}$ denoting light speed in vacuum is a constant, the relativistic diffusion equation  reduces to 
\begin{equation}
 \nabla ^2 \gamma =0.
\end{equation}
The physical meaning is that there is no entropy production for light transport in vacuum.

The rest terms in Eq.~\eqref{eq:20}  is the relativistic kinetic Gibbs-Duhem relation:
\begin{equation}
 \mathbf c\cdot  \mathrm d (\gamma m_0\mathbf c)+ \mathrm d P=0.
\end{equation}
Integrating the above equation results in an explicit expression for the 
relativistic pressure:
\begin{equation}
 P=-\gamma m_0 \mathbf{c}^2+C.
\end{equation}
Using the condition that $P=0$ when $\mathbf u=\mathbf 0$, i.e., $\gamma=1$, we obtain the expression for the integration constant: $C=m_0\mathbf{c}^2$.
Therefore, the relativistic pressure is expressed as 
\begin{equation}
 P=(1-\gamma) m_0 \mathbf{c}^2.
\end{equation}
Thus, the effective system energy is expressed as 
\begin{equation}
 E=m_0\mathbf{c}^2.
\end{equation}
In the idealized case, the integration constant vanishes by imposing the condition \(P=0\) when \(|\mathbf{u}|=0\). In contrast, the realistic case generally permits a nonzero background energy, even in the absence of fluid motion. This residual energy may provide a possible interpretation of nonzero vacuum energy~\cite{scali2025cosmological,ferrero2025quantum,epelbaum2024vacuum}.

\section{Discussion, validation and new interpretation}

\subsection{Linear momentum and its validation}

Conventionally, the linear momentum balance equation is derived from Noether's theorem through the invariance of the underlying physical system under spatial translations~\cite{noether1971}. In the present work, we propose an alternative interpretation based on the kinetic Gibbs--Duhem relation. Specifically, we derive the linear momentum balance as a consequence of the constraint imposed by Galilean invariance of the system kinetic energy. From this perspective, the momentum balance equation is not the fundamental requirement for Galilean invariance; rather, it emerges as a consequence of the Galilean invariance of the system energy.

Within the proposed framework, the kinetic Gibbs--Duhem relation further establishes a connection between pressure and microscopic or macroscopic fluid motion. Specifically, the hydrostatic pressure is associated with microscopic particle motion, whereas the hydrodynamic pressure is associated with macroscopic fluid motion. In the latter case, the Galilean-invariant relation
\begin{equation}
    p+\rho_{\mathrm{macro}}\mathbf{u}_{\mathrm{macro}}^2=C
\end{equation}
recovers the Bernoulli relation. In the former case, the corresponding relation
\begin{equation}
    p+\rho_{\mathrm{micro}}\mathbf{u}_{\mathrm{micro}}^2=C
\end{equation}
recovers the ideal-gas equation of state within the proposed framework. These results were examined in detail in the first part of this research series~\cite{wang2026density} and are not repeated here for brevity.

For macroscopic fluid motion, we introduce the relation for the diffusivity $D_1$ as 
\begin{equation}
    D_1=\frac{\eta}{m_1^2},
    \label{eq:28-0}
\end{equation}
where $\eta$ denotes the dynamic viscosity. The velocity diffusion equation can then be written as
\begin{equation}
    \frac{\mathrm{d}\mathbf{u}}{\mathrm{d}t}
    =\nabla\cdot\left[
    \frac{\eta}{m_1^2}\nabla(m_1\mathbf{u})
    \right].
\end{equation}
For constant $m_1$ and $\eta$, this equation reduces to
\begin{equation}
    m_1\frac{\mathrm{d}\mathbf{u}}{\mathrm{d}t}
    =\eta\nabla^2\mathbf{u}.
    \label{eq:29}
\end{equation}
The right-hand side has the same mathematical form as the divergence of the Newtonian viscous stress for an incompressible fluid with constant viscosity, where the viscous stress tensor is expressed as $\boldsymbol{\tau}=\eta\nabla\mathbf{u}$ under the present simplified formulation.

Comparing Eq.~\eqref{eq:29} with the momentum evolution equation [Eq.~\eqref{eq:12}], namely,
\begin{equation}
    m_1\frac{\mathrm{d}\mathbf{u}}{\mathrm{d}t}
    =-\nabla P_1,
\end{equation}
we obtain
\begin{equation}
    \eta\nabla^2\mathbf{u}+\nabla P_1=0.
\end{equation}
This equation takes the form of the classical Stokes momentum equation for incompressible viscous flow, thereby establishing a connection between the proposed velocity diffusion framework and conventional viscous fluid dynamics.

\subsection{Non-linear momentum and its quantum interpretation}

In the Eulerian framework, neglecting convection, the second-order velocity diffusion equation [Eq.~\eqref{eq:15-9}] is given by
\begin{equation}
    \frac{\partial \mathbf{u}^2}{\partial t}
    =D_2\nabla^2(m_2\mathbf{u}^2)
    =\frac{D_2m_0}{c^2}\nabla^2\mathbf{u}^2,
\end{equation}
where \(m_2=m_0/c^2\), with the binomial coefficient omitted for brevity. Note that the effective mass parameter \(m_2\) is scaled by the inverse square of the speed of light and is therefore small in the present formulation. At this effective mass scale, a deterministic particle-trajectory description may be insufficient to characterize the microscopic state of the system. Instead, the particle's spatial distribution is described statistically, motivating a probabilistic description of its position. Within a quantum-mechanical interpretation, this spatial probability distribution can be represented by a wave function. Accordingly, we introduce an effective wave function by rescaling \(\mathbf{u}^2\):
$\psi=A\mathbf{u}^2$, where \(A\) is a scalar scaling parameter.

When higher-order terms are considered, this definition can be generalized to a vector-valued wave function of the form
\[
\boldsymbol{\psi}
=\left(A_1\mathbf{u}^2,\; A_2\mathbf{u}^4,\; \ldots\right),
\]
where \(A_1,A_2,\ldots\) are scalar scaling parameters. In the following discussion, we focus on the scalar wave function.

Similar to Eq.~\eqref{eq:28-0}, we define the diffusion coefficient as
\begin{equation}
    D_2=i\frac{\hbar c^2}{2m_0^2},
    \label{eq:34}
\end{equation}
which allows the second-order velocity diffusion equation to be recast in the form of a Schrödinger equation for fluids:
\begin{equation}
    i\hbar\frac{\partial\psi}{\partial t}
    =-\frac{\hbar^2}{2m_0}\nabla^2\psi.
\end{equation}

Comparing Eq.~\eqref{eq:34} with Eq.~\eqref{eq:28-0} suggests a new physical interpretation of microscopic viscosity: within the proposed framework, the effective viscosity in the second order velocity dissipation
at the microscopic scale is associated with the reduced Planck constant, i.e.,
\begin{equation}
 \eta^*=\frac{1}{2}\hbar c^2,\quad  \text{so that} \quad D_2=i\frac{\eta^*}{m_0^2},
\end{equation}
The imaginary unit in the diffusion coefficient accounts for the complex-valued evolution of the wave function. Alternatively, the evolution can be formulated in terms of real-valued equations by separating the wave function into its real and imaginary parts. In that representation, the imaginary unit does not explicitly appear in the resulting coupled equations.

For a plane-wave solution of the form
$\psi=\psi_0\exp(ikx+\omega t)$,
where \(k\) is the wave number and \(\omega\) is the temporal growth or decay rate, the Laplacian of the wave function is given by \(\nabla^2\psi=-k^2\psi\). Substituting this relation into the Schrödinger equation yields
\begin{equation}
    i\frac{\hbar}{c}\frac{\partial\psi}{\partial t}
    =\frac{\hbar^2 k^2}{2m_0c}\psi
    =\beta m_0 c\,\psi,
    \label{eq:30}
\end{equation}
where the dimensionless coefficient \(\beta\) is defined as
\begin{equation}
    \beta=\frac{1}{2} \bigg(\frac{\hbar k}{m_0 c}\bigg)^2=\frac{1}{2} \bigg(\frac{u^2}{ c^2}\bigg).
\end{equation}

Furthermore, noting that \(P_2=-m_2\mathbf{u}^4=m_2\psi^2/A^2\), the second-order momentum evolution equation [Eq.~\eqref{eq:19-9}] neglecting convection in the Eulerian framework can be expressed as
\begin{equation}
    \mathbf{u}\frac{\partial\psi}{\partial t}
    =-\frac{1}{m_2}\nabla P_2
    =\frac{1}{A^2}\nabla\psi^2.
\end{equation}
Taking the dot product of both sides with the velocity vector \(\mathbf{u}\) yields
\begin{equation}
    u^2\frac{\partial\psi}{\partial t}
    =\frac{2\psi}{A^2}\mathbf{u}\cdot\nabla\psi.
\end{equation}
Consequently, the resulting scalar equation can be recast in the form
\begin{equation}
    i\frac{\hbar}{c}\frac{\partial\psi}{\partial t}
    =i\hbar\boldsymbol{\alpha}\cdot\nabla\psi,
    \label{eq:32}
\end{equation}
where
\begin{equation}
    \boldsymbol{\alpha}
    =\frac{2\psi}{A^2\mathbf{u}^2}\frac{\mathbf{u}}{c}
\end{equation}
is a vector-valued function of \(\psi\) and \(\mathbf{u}\).

Combining Eqs.~\eqref{eq:30} and \eqref{eq:32}, we obtain a Dirac-type equation for the fluid:
\begin{equation}
    i\frac{\hbar}{c}\frac{\partial\psi}{\partial t}
    =\left(i\hbar\boldsymbol{\alpha}\cdot\nabla
    +\beta m_0 c\right)\psi.
\end{equation}
This Dirac-type equation combines the contributions from velocity diffusion and momentum evolution in the second-order expansion of the kinetic energy. Within the proposed framework, it provides a unified description of the associated wave-function and momentum dynamics.

\section{Conclusion}

In summary, we have presented an alternative theoretical framework for interpreting the physical origin of nonlinear momentum evolution in fluid dynamics. In particular, our analysis suggests that relativistic effects can give rise to nonlinear momentum evolution even in the low-velocity limit, offering a different perspective on the mechanisms governing fluid motion.

Furthermore, by interpreting the second-order velocity diffusion equation within a quantum-mechanical framework, we have established a connection between higher-order viscous transport and the reduced Planck constant. Specifically, we identify the second-order viscosity coefficient as
\begin{equation}
    \eta^*=\frac{1}{2}\hbar c^2.
\end{equation}
This result suggests a potential link between higher-order viscous transport and quantum dynamics, providing an alternative interpretation of second-order velocity diffusion.

Finally, we have shown that, within the proposed framework, the coupling between second-order velocity diffusion and nonlinear momentum evolution leads to a Dirac-type equation for quantum fluids in the low-fluid-velocity regime. This result establishes a theoretical connection among relativistic fluid dynamics, higher-order viscous transport, and quantum wave mechanics.

Overall, the present framework offers an alternative perspective on nonlinear momentum evolution in fluids and highlights the potential role of relativistic and quantum effects in the emergence of higher-order fluid dynamics.

\acknowledgments

I am grateful to R. Yin for many valuable discussions.

\end{document}